\documentclass[a4, showpacs, notitlepage, nofootinbib]{elsarticle}
\usepackage{graphicx}
\usepackage[siunitx]{circuitikz}
\usepackage{amsmath}
\usepackage{filecontents, url}
\usepackage{physics}
\usepackage{braket}
\usepackage{subcaption}
\usepackage{lipsum}
\usepackage{natbib}
\usepackage{mathtools}
\usepackage{footnote}
\usepackage{textcomp}
\usepackage{eqnarray,amsmath}
\usepackage{geometry}

\begin{document}

\title{An investigation of higher order moments of empirical financial data and the implications to risk}
\author{Luke De Clerk}
\address{Department of Physics, Loughborough University, Leicestershire, LE11 3TU, United Kingdom}
\ead{l.de-clerk@lboro.ac.uk}
\author{Sergey Savel'ev}
\address{Department of Physics, Loughborough University, Leicestershire, LE11 3TU, United Kingdom}
\ead{s.saveliev@lboro.ac.uk}
\date{\today}

\begin{abstract}
Here, we analyse the behaviour of the higher order standardised moments of financial time series when we truncate a large data set into smaller and smaller subsets, referred to below as time windows. We look at the effect of the economic environment on the behaviour of higher order moments in these time windows. We observe two different scaling relations of higher order moments when the data sub sets' length decreases; one for longer time windows and another for the shorter time windows. These scaling relations drastically change when the time window encompasses a financial crisis. We also observe a qualitative change of higher order standardised moments compared to the gaussian values in response to a shrinking time window. We extend this analysis to incorporate the effects these scaling relations have upon risk. We decompose the return series within these time windows and carry out a Value-at-Risk calculation. In doing so, we observe the manifestation of the scaling relations through the change in the Value-at-Risk level. Moreover, we model the observed scaling laws by analysing the hierarchy of rare events on higher order moments.
\end{abstract}

\begin{keyword}
Empirical data, Scaling relations, higher order standardised moments, Value-at-Risk, Gaussian Mixtures
\end{keyword}

\maketitle
\textit{JEL Classification: C10, G01}
\section{Introduction}
In many financial settings, the behaviour of market data is analysed to better understand: the logarithmic price change, \cite{engle_ap, go_garch, asym1, nelson_garch}, the historic or implied volatility, \cite{engle_real, garch_optionpricing} or the actual price behaviour \cite{BSEqt, asym4, PP3}. Nevertheless, in \cite{me_1}, the higher order moments were used to study the applicability of certain Generalised AutoRegressive Conditional Heteroskedasticity (GARCH) models for mimicking price dynamics. The use of higher order moments within financial modelling is well established, \cite{4}. By investigating the higher order moments we can get an insight to the distribution of price change and how it varies over time. By doing this, we can evaluate the hypothesis that rare-events originate from huge volatility shocks, as such this phenomena is likely seldom seen in short time windows and is much more likely in long time windows. This observation helps us to understand behaviour of higher order central moments in different time windows.

The higher order moments are used in this investigation due to their ability to capture the general aspects of the distribution of price change, \cite{high_mom1, high_mom2, high_mom3}. The higher order moments show the quantity of outliers within the distribution, \cite{4}. If the fourth order standardised statistical moment (also called kurtosis) of empirical data sets are larger than 3, we have a leptokurtic distribution. This manifests itself in a larger probability of getting an outcome that is much larger or smaller than the mean. Such a behaviour is also known as a rare-event. Therefore, we can study the properties of the time series without the need for many different metrics.

We will consider time windows of $N$ trading days and analyse the logarithm of stock returns defined as:
\begin{equation}
x_i = \ln \left( \frac{y(t_0 + i\delta t)}{y(t_0+(i-1)\delta t)} \right)
\end{equation}
where $t_0$ being the date of the first trading day within the studied window, $y(t_0+(i-1)\delta t)$ is the closing price on the $i$th trading day with $1\leq i \leq N$ and $\delta t$ referring to the time between trading days. Using the logarithm of stock returns, we estimate the nth order standardised moments:
\begin{equation}
\Gamma_n(t_0,N)  = \frac{\langle (x-\mu)^n \rangle}{\langle (x-\mu)^2 \rangle^{\frac{n}{2}}}
\label{nth}
\end{equation}
where,
\begin{equation}
\langle (x-\mu)^n\rangle = \frac{1}{N}\sum_{i=1}^N (x_i-\mu)^n
\end{equation}
and,
\begin{equation}
\mu = \frac{1}{N} \sum_{i=1}^N x_i.
\end{equation}
We analyse the dependence of $\Gamma_n(t_0,N)$ for $n=4$ and $n=6$ on both number of trading days $N$ in the window and the absolute time $t_0$ with a goal to observe and analyse empirical laws and their evolution during the economic crisis. In addition, we compare $\Gamma_n(t_0,N)$ for different $N$ and $n=2, 4, 6, 8, 10, 12$ with the corresponding values of gaussian standardised moments.

The paper is organised as follows; in section \ref{scale}, we introduce the scaling relations in ($\Gamma_4$, $\Gamma_6$) space. In section \ref{trunc}, we introduce the economic periods we wish to analyse whilst presenting the results for the empirical data. In Section \ref{kurt_emp}, we compare the higher order standardised moments obtained from the empirical data with the corresponding moments for the gaussian distribution. Section \ref{sim} shows that scaling relations are linked to the hierarchy of rare events and the exponent of price power law distributions. In addition, we develop an approach linking the obtained scaling with stock risk management inspired by previous studies, \cite{capm1, bus_cycles, conf_1, conf_2}, where higher order moments were used for risk assessment. In section \ref{risk}, we highlight the change in behaviour for the Value-at-Risk for the same time windows when we observe changes of the scaling relations of higher order moments. Finally, section \ref{conc} concludes.

\section{Empirical Data Processing}

\subsection{Scaling relations in ($\Gamma_4$, $\Gamma_6$) space with increasing time windows of averaging}
\label{scale}
We truncate an 18 year (6th October 2000 to 6th October 2018) time series into a $1\%$ time window, and then we gradually increase the length of the time window up to $100\%$ of the whole time series, corresponding to 4536 days.

\begin{figure}[h!]
\includegraphics[width=\textwidth]{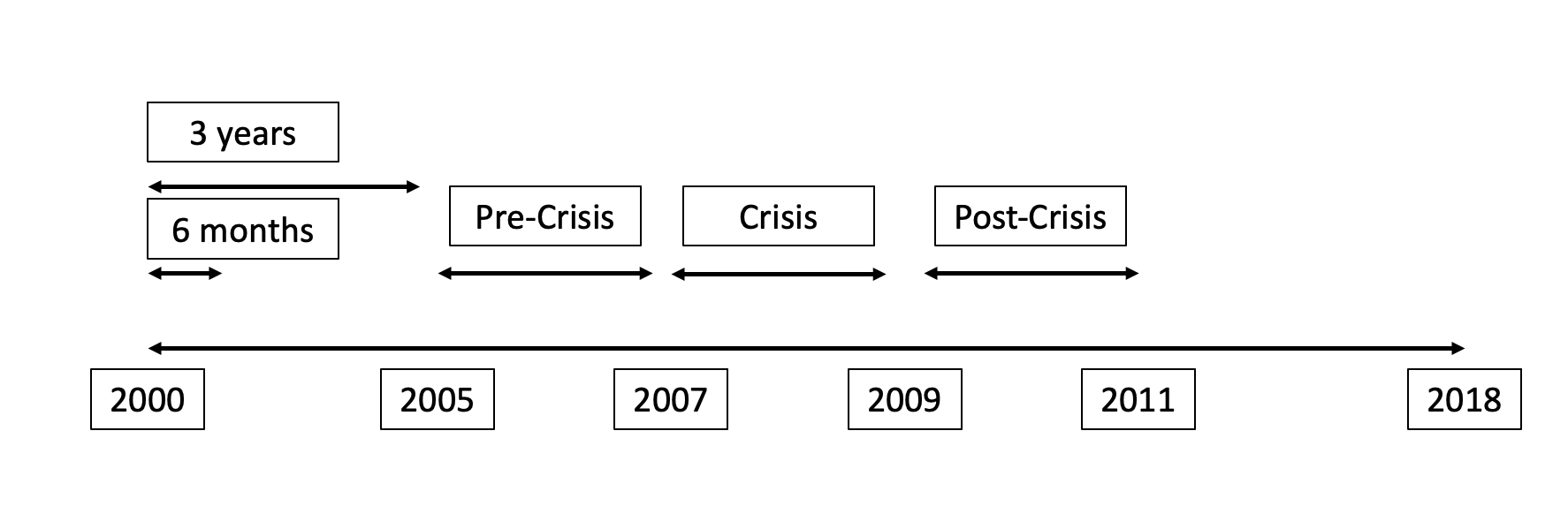}
\caption{The timeline of time series used for the three periods of economic cycles and the truncation data, \cite{me_1}.}
\label{timeline}
\end{figure}

Calculating the fourth and sixth order statistical moments of the empirical data in all windows (see the method described above), we are able to present the data in ($\Gamma_4$, $\Gamma_6$) space. To analyse the behaviour of the market data in response to the truncation of the time series we propose to use scaling relations:

\begin{equation}
\Gamma_6 = A\Gamma_4^B
\label{scale_eqt}
\end{equation}
where A and B are constants. In logarithmic scale this reduces to a straight line:

\begin{equation}
\ln(\Gamma_6) = B\ln(\Gamma_4) + \ln(A).
\end{equation}
Below, we will refer to the parameter $B$ as either the scaling exponent or the logarithmic gradient.

 \begin{figure*}
        \centering
        \begin{subfigure}[b]{0.475\textwidth}
            \centering
            \includegraphics[width=\textwidth]{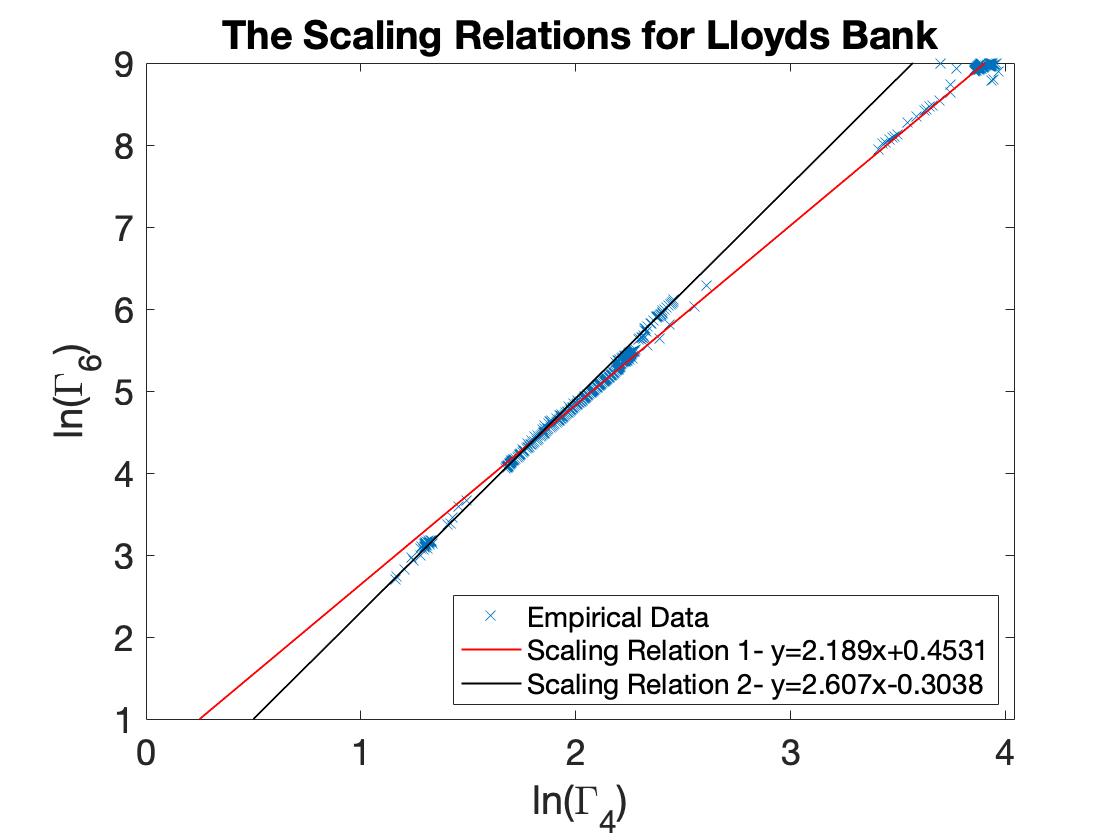}
            \caption[]%
            {{\small}}    
            \label{lloyds_truncated}
        \end{subfigure}
        \hfill
        \begin{subfigure}[b]{0.475\textwidth}  
            \centering 
            \includegraphics[width=\textwidth]{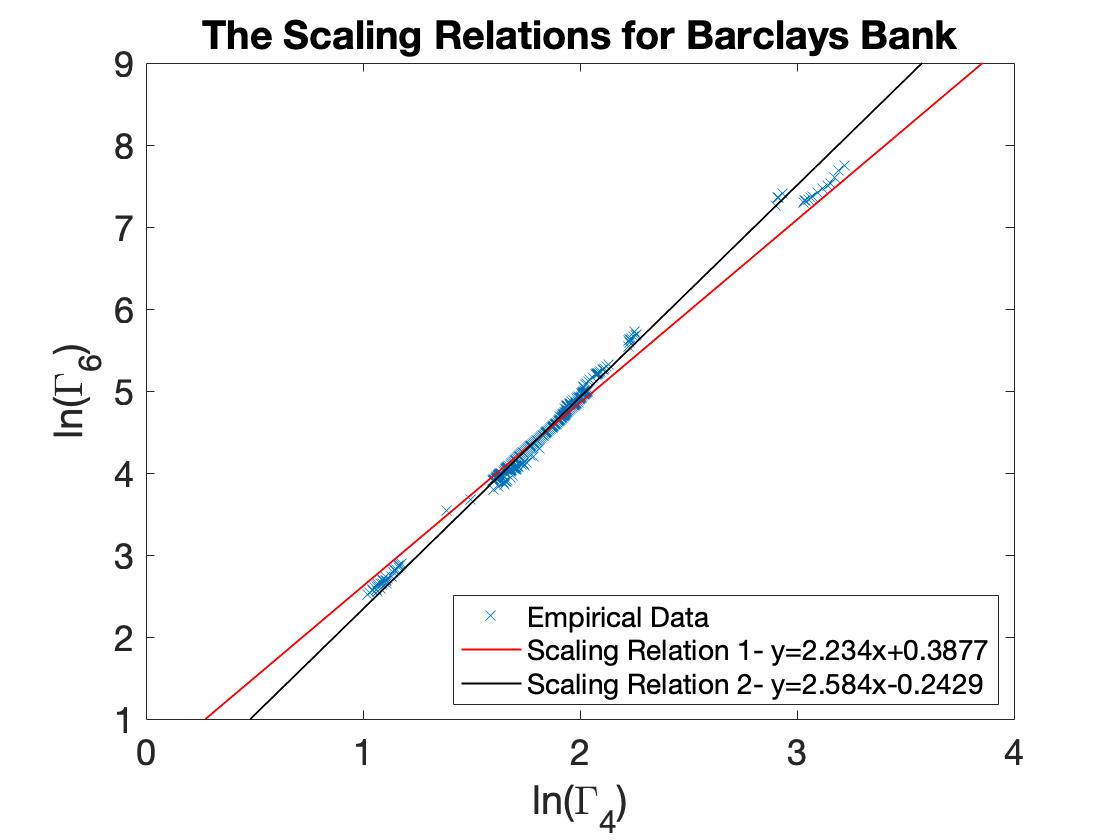}
            \caption[]%
            {{\small}}    
            \label{barc_truncated}
        \end{subfigure}
        \vskip\baselineskip
        \begin{subfigure}[b]{0.475\textwidth}   
            \centering 
            \includegraphics[width=\textwidth]{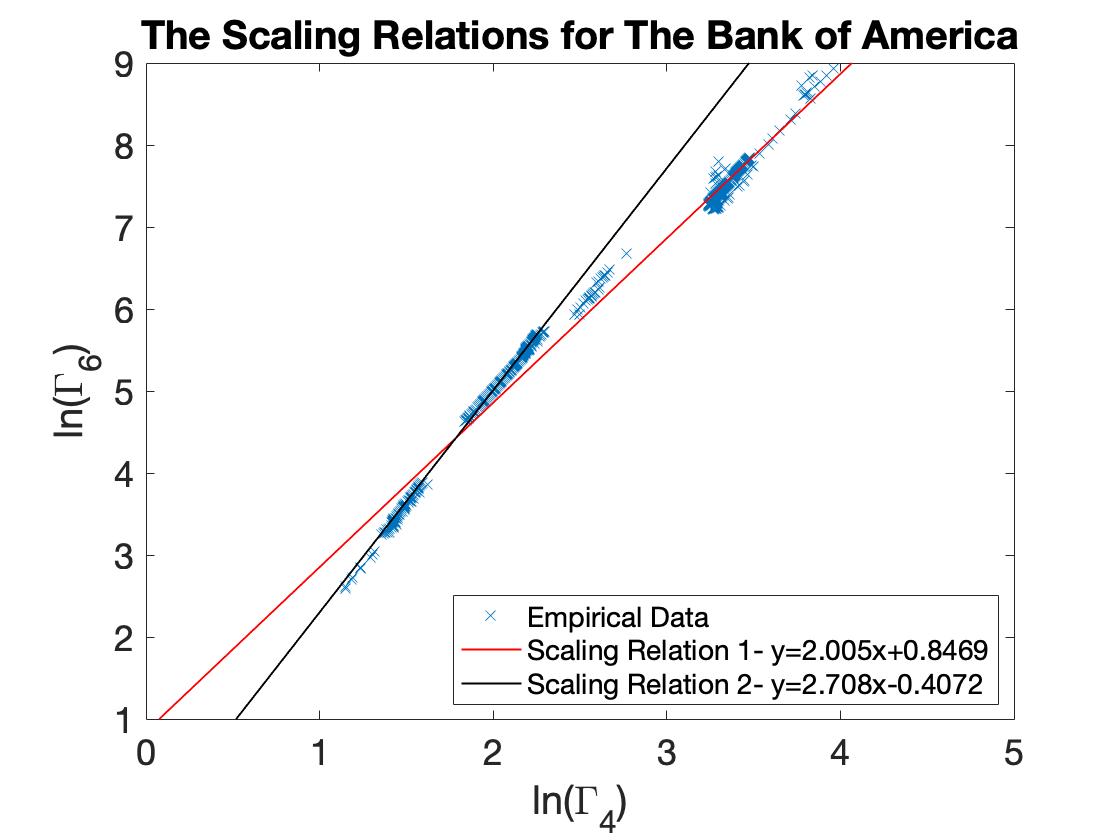}
            \caption[]%
            {{\small}}    
            \label{boa_truncated}
        \end{subfigure}
        \hfill
        \begin{subfigure}[b]{0.475\textwidth}   
            \centering 
            \includegraphics[width=\textwidth]{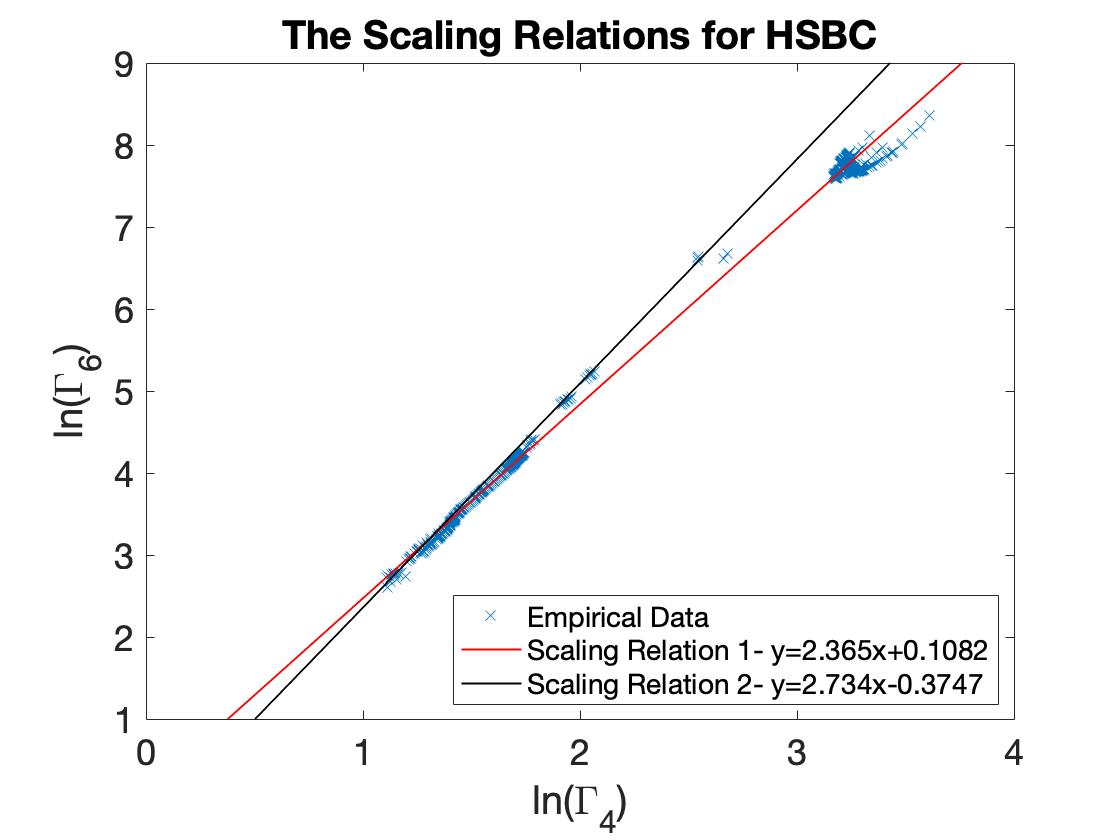}
            \caption[]%
            {{\small}}    
            \label{hsbc_truncated}
        \end{subfigure}
        \caption[]
        {\small Here, we show the data points on the ($\ln(\Gamma_4)$, $\ln(\Gamma_6)$) phase space and the two scaling relations discussed in the text. The black line shows the scaling relation for the shorter time windows, whilst the red line shows the scaling relation for the longer time windows. In panel (a), we show Lloyds Bank, (b), Barclays Bank, (c), Bank of America and (d), HSBC.} 
        \label{truncated}
 \end{figure*}

Figure \ref{truncated} uncovers two different scalings for four banking companies. The first relation, shown in red for all companies, is the scaling relations for the longer time windows. However, we discover a different scaling behaviour with different exponents $B$ for shorter time windows. Such a distinct two scaling behaviour has been observed for all companies studied. We also observe similar scaling relations for other even standardised moments. For example, in appendix \ref{g8_scale}, we present the scaling relations for the fourth $\Gamma_4$ and eighth $\Gamma_8$ standardised moments.

\subsection{The Impact of Economic Environments upon Scaling Relations}
\label{trunc}
We now turn our attention to the effect of the economic environment on the scaling relations. To investigate this we use the economic periods set out in figure \ref{timeline}. Here, we have a pre-crisis period, 2005, before the financial crash, a crisis period, 2008 and then the post-crisis period, 2011. For completeness, we investigate the succeeding years of 2014 and 2017, to see the effect the financial crisis has upon the scaling relations over a prolonged period of time.

\begin{table}[h!]
\centering
\begin{tabular}[c]{|p{1.5cm}|p{1cm}|p{1cm}|p{1cm}|p{1cm}|p{1cm}|}
\hline
Company&2005&2008&2011&2014&2017\\ \hline
Barclays&${\cal Y}=7.4{\cal X}-9.8$&${\cal Y}=31.3{\cal X}-112.4$&${\cal Y}=12.1{\cal X}-25.4$&${\cal Y}=12.6{\cal X}-23.8$&${\cal Y}=8.3{\cal X}-12.4$\\ \hline
Bank Of America&${\cal Y}=6.7{\cal X}-8.5$&${\cal Y}=18.7{\cal X}-45.4$&${\cal Y}=24.8{\cal X}-66.4$&${\cal Y}=9.6{\cal X}-14.5$&${\cal Y}=15.7{\cal X}-29.8$\\ \hline
Gold&${\cal Y}=9.3{\cal X}-15$&${\cal Y}=28.3{\cal X}-87.2$&${\cal Y}=10.5{\cal X}-15.9$&${\cal Y}=23.4{\cal X}-68.4$&${\cal Y}=12.4{\cal X}-25$\\ \hline
GSK&${\cal Y}=6.4{\cal X}-3$&${\cal Y}=11.7{\cal X}-20.2$&${\cal Y}=13.2{\cal X}-25.8$&${\cal Y}=33{\cal X}-70.6$&${\cal Y}= 22.12{\cal X}-49.9$\\ \hline
Lloyds&${\cal Y}=21.2{\cal X}-44$&${\cal Y}=35.2{\cal X}-172.9$&${\cal Y}=12.3{\cal X}-23.2$&${\cal Y}=11.4{\cal X}-18.2$&${\cal Y}=8.9{\cal X}-11$\\ \hline
Rio Tinto&${\cal Y}=9.8{\cal X}-14.8$&${\cal Y}=14{\cal X}-27.1$&${\cal Y}=10.7{\cal X}-19$&${\cal Y}=9.3{\cal X}-14.6$&${\cal Y}=11{\cal X}-17$\\ \hline
\end{tabular}
\caption{Here, we present the scaling relations for the longer time horizons for several companies of different economic environments. ${\cal Y}$, represents the logarithm of the sixth order standardised moment and ${\cal X}$ the logarithm of the fourth order standardised moment, with the coefficient in front of ${\cal X}$ being the logarithmic gradient, $B$.}
\label{table_scaling}
\end{table}

The scaling relations for these periods are worked out using the same method as described above for a 252 day time series, however, we do not consider time windows shorter than 25 days, that is we ignore time windows with very small statistics. The results can be found in table \ref{table_scaling}, where ${\cal Y}$ is the $ln(\Gamma_6)$ and ${\cal X}$ is the $\ln(\Gamma_4)$ for the longer window scaling. It is clear from the scaling relations found, the economic period has a very vivid effect upon the companies behaviour. 

For instance, if we analyse Lloyds Bank. The scaling relation for the pre-crisis period, has a logarithmic gradient $B=21.2$, whereas, within and just after the crisis period the exponent, $B$ increases drastically. After the crisis period, the exponent, $B$, decreases to a lower level than the pre-crisis period. This indicates, the long term impact of the financial crash. As we do not see this behaviour of the exponent in other security types, we can infer that this behaviour is due to these companies being directly affected by the financial crash of 2008. The fact we have persistence of this effect can be seen as an indication that the financial crisis period has long run dynamical impacts upon the market price of these companies. The same behaviour, however, can be seen for the Gold scaling relations. We see an increase of the exponent, $B$, in response to the financial crisis, followed by the exponent returning to a level similar to the pre-crisis environment. However, the striking increase of logarithmic gradient, $B$, followed by its post-crisis drop observed for banking companies, has not been found in non-banking sectors of the economy. For example, GSK (a pharmaceutical company) and Rio Tinto (a metals and mining corporation) do not have such a distinguished behaviour. In the case of Rio Tinto, the gradient stays relatively static throughout the time, whereas, GSK has an increase in 2014, which could be attributed to the bribery scandal that encompassed the company from 2013 to 2014, \cite{gsk}. It can therefore be said that the financial crisis has more of an effect upon the banking companies, as we would expect due to the nature of the cause of the crisis period.

\subsection{Higher Order Standardised Moments in Empirical Data}
\label{kurt_emp}
Here, we compare the higher order standardised moments defined by equation (\ref{nth}) for $n=$4, 6, 8, 10 and 12 and compare them with the corresponding gaussian standardised moment values listed in table \ref{kurt_ratios}. The equation for this ratio, $R_n$, is shown below: 

\begin{equation}
R_n = \frac{\Gamma_n^{gaussian}}{\Gamma_n}.
\label{gauss_ratio}
\end{equation}
The results for various companies and their market data can be seen in figure \ref{kurt_ratios}. We also show ratio (\ref{gauss_ratio}), for shorter time windows, namely, 3 years and 6 months. The results of which can be seen in figure \ref{short_kurt}. The values of the higher order standardised moments of the empirical financial series for the 3 years time window can be found in table \ref{kurt_short_table}, in appendix \ref{kurt}.

\begin{figure}[h!]
\centering
\includegraphics[width=0.75\linewidth]{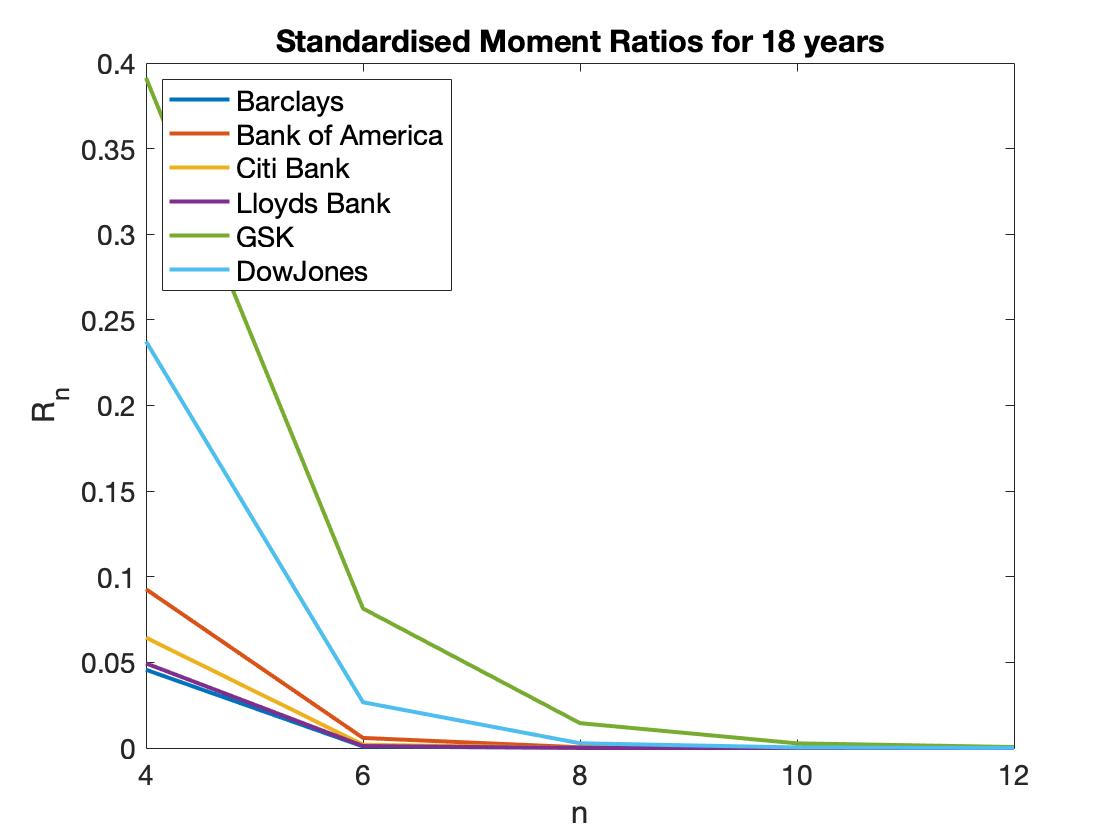} 
\caption{The standardised moment ratios for varying orders, $n$, see appendix \ref{kurt}, for the gaussian standardised moments divided by the empirical values. Here, we have analysed the 18 year time series for Bank of America, Barclays Bank, Citi Bank, the DowJones Index, GSK, HSBC and Lloyds Bank.}
\label{kurt_ratios}
\end{figure}

\begin{figure}[h!]
\centering
\begin{subfigure}{0.5\linewidth}
  \centering
  \includegraphics[width=\linewidth]{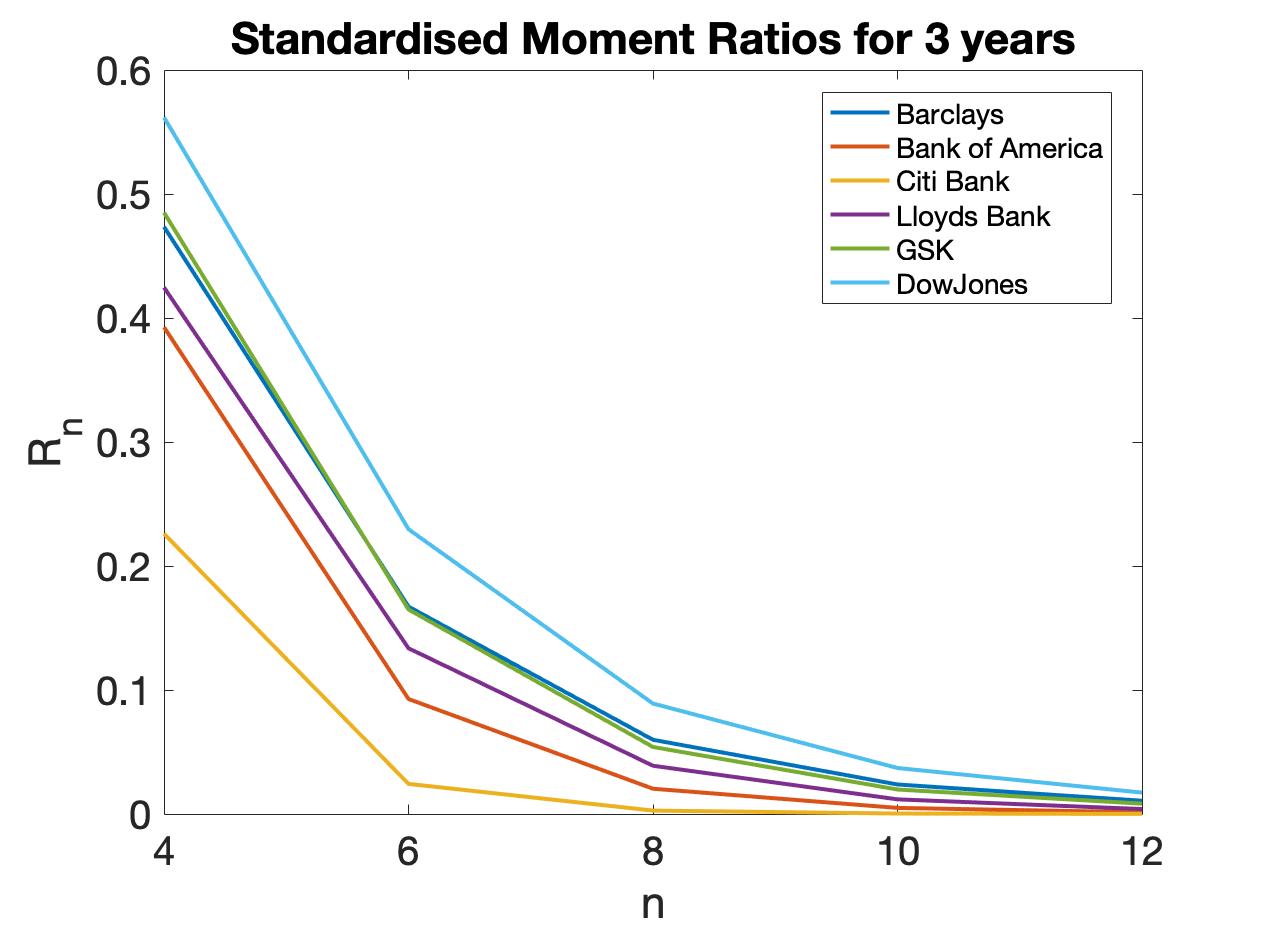}
  \caption{}
  \label{3year}
\end{subfigure}%
\begin{subfigure}{0.5\linewidth}
  \centering
  \includegraphics[width=\linewidth]{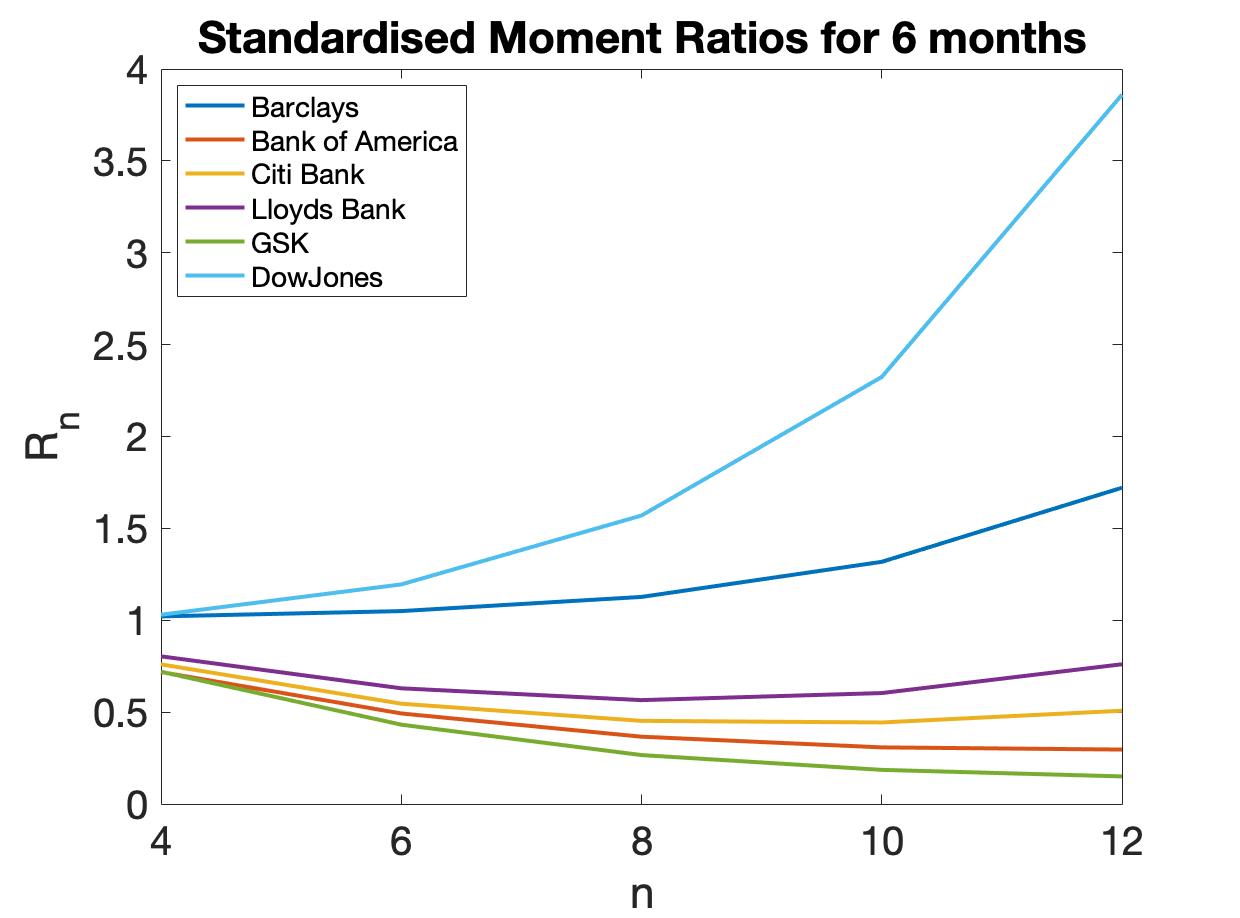}
  \caption{}
  \label{6month}
\end{subfigure}
\caption{The ratios for the gaussian higher order standardised moments to the empirical ones. Here we have analysed, the 3 year, panel (a), and the 6 month, panel (b), time series for the Bank of America, Barclays Bank, Citi Bank, the DowJones Index, GSK and Lloyds Bank, for 6th October 2015-6th October 2018, panel (a), and 6th April 2018-6th October 2018, panel (b). Remarkably, for some companies' 6 month time series the ratio of gaussian to empirical standardised moments is above one in contrast to results shown in figure \ref{kurt_ratios} and \ref{3year}. In addition, Lloyds Bank (purple curve) shows a non-monotonic behaviour as standardised moment order increases.}
\label{short_kurt}
\end{figure}

The evolution of higher order moments of the empirical financial series is quite remarkable. When we take a long time series, either the 18 years or the 3 years (figures \ref{kurt_ratios} or \ref{3year}), the ratio of gaussian to empirical standardised moment is below 1, which we can expect for leptokurtic distributions. When we instead truncate this time series to 6 months, figure \ref{6month}, we get some empirical higher order standardised moments that are now less than that of the gaussian values. Moreover, we uncover the decay of the ratios as a function of its order for long series (18 and 3 years) which unexpectedly start to grow or even have a non-monotonic behaviour for shorter time windows (6 months).

\section{GARCH-double-normal Simulations}
\label{sim}
To understand the origins of these different scaling relations we simulate a GARCH-double-normal(1,1) model as seen in, \cite{me_1, dg_1, dg_2}. The dynamic equation for $\sigma_t^2$, the conditional variance, is given by:

\begin{equation}
\sigma_t^2 = \alpha_0 + \alpha_1 x_{t-1}^2 + \beta_1 \sigma_{t-1}^2
\end{equation}
where, $x_t = \chi_t \sigma_t$, and $\chi_t$ is an independent identically distributed random variable with standard deviation equal to 1. For the Double Gaussian GARCH model, the distribution of $\chi_t$ has the form:

\begin{equation}
p(x) = \frac{a}{\sigma_1\sqrt{2\pi}}e^{-x^2/(2\sigma_1^2)} + \frac{b}{\sigma_2\sqrt{2\pi}}e^{-x^2/(2\sigma_2^2)}
\label{dg_dist}
\end{equation}
We assign the following values for the parameters of the distribution (\ref{dg_dist}); $a=0.9818$, $b=0.0182$, $\sigma_1^2 = 0.833$ and $\sigma^2_2 = 9.986$. We simulate, a return time series and truncate the series in the exact manner that has been undertaken for the empirical data above. To do this, we use the following GARCH parameter values; $\alpha_0 = 1e-5$, $\alpha_1 = 0.5$ and $\beta_1 = 0$.

\begin{figure}[h!]
\centering
\includegraphics[width=0.75\linewidth]{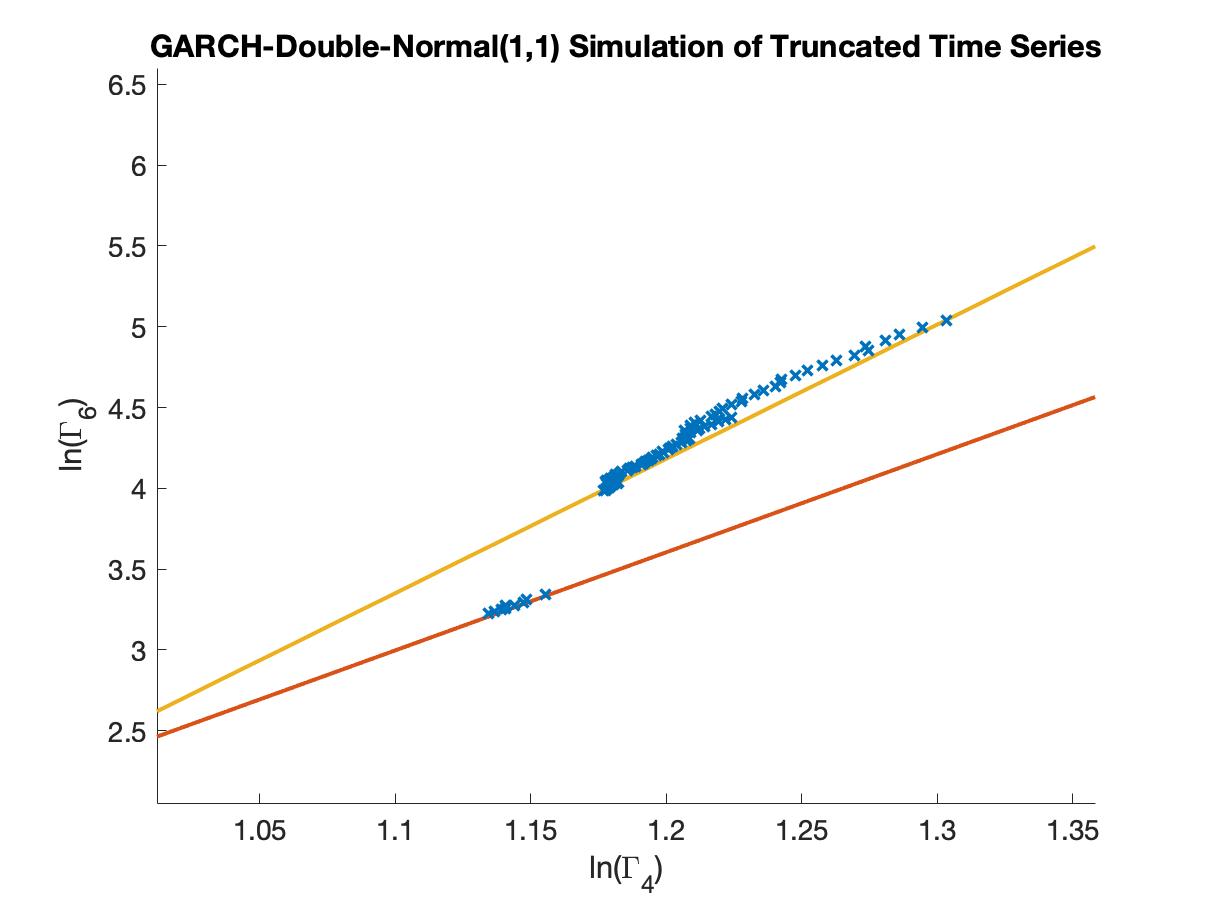} 
\caption{Two distinct scaling relations are evident in the fourth and sixth order standardised higher order moments obtained from the simulation data of a GARCH-double-normal(1,1) model. Here we see a scaling relation for the shorter time windows, shown by the red line, given by the equation, $y = 6.07x - 3.68$ and one for the longer time windows, shown by the yellow line, given by the equation, $y = 8.31x-5.79$.} 
\label{dg_scale}
\end{figure}

The results for the GARCH simulation of studied scaling laws can be seen in figure \ref{dg_scale}. Here, we see the two scaling relations for the different time scales. For the shorter time windows (the red line), we have a straight line equation of; $y = 6.07x - 3.68$ and for the longer time horizons (the yellow line), we have the equation; $y = 8.31x - 5.79$. There is a clear difference in the scaling relations of the simulated data and the empirical data, namely in the simulated data the logarithmic gradient, $B$, for the shorter time window is lower than the longer time window. Something that is not mirrored in the empirical data. Now, if we change the parameters $\alpha_1$ and $\beta_1$ we will be able to highlight how the dependence on the past level of return and past level of volatility impacts the scaling relations. If we increase $\beta_1$, to 0.8 and reduce $\alpha_1$ to 0.1, we see that the longer time horizon scaling relation is still steeper than the shorter one. The same is true when we increase $\alpha_1$ to 0.6 and reduce $\beta_1$ to 0.1. However, we do see a connection of the value of $\alpha_1$ to the value of the logarithmic gradient for the scaling relations. When we increase $\alpha_1$, the logarithmic gradient, $B$, also increases. That is to say, the more of a dependence the past return has on the future volatility level, the larger the value of the logarithmic gradient, $B$.

\section{Implications on Value-at-Risk}
\label{risk}
\begin{figure*}
        \centering
        \begin{subfigure}[b]{0.475\textwidth}
            \centering
            \includegraphics[width=\textwidth]{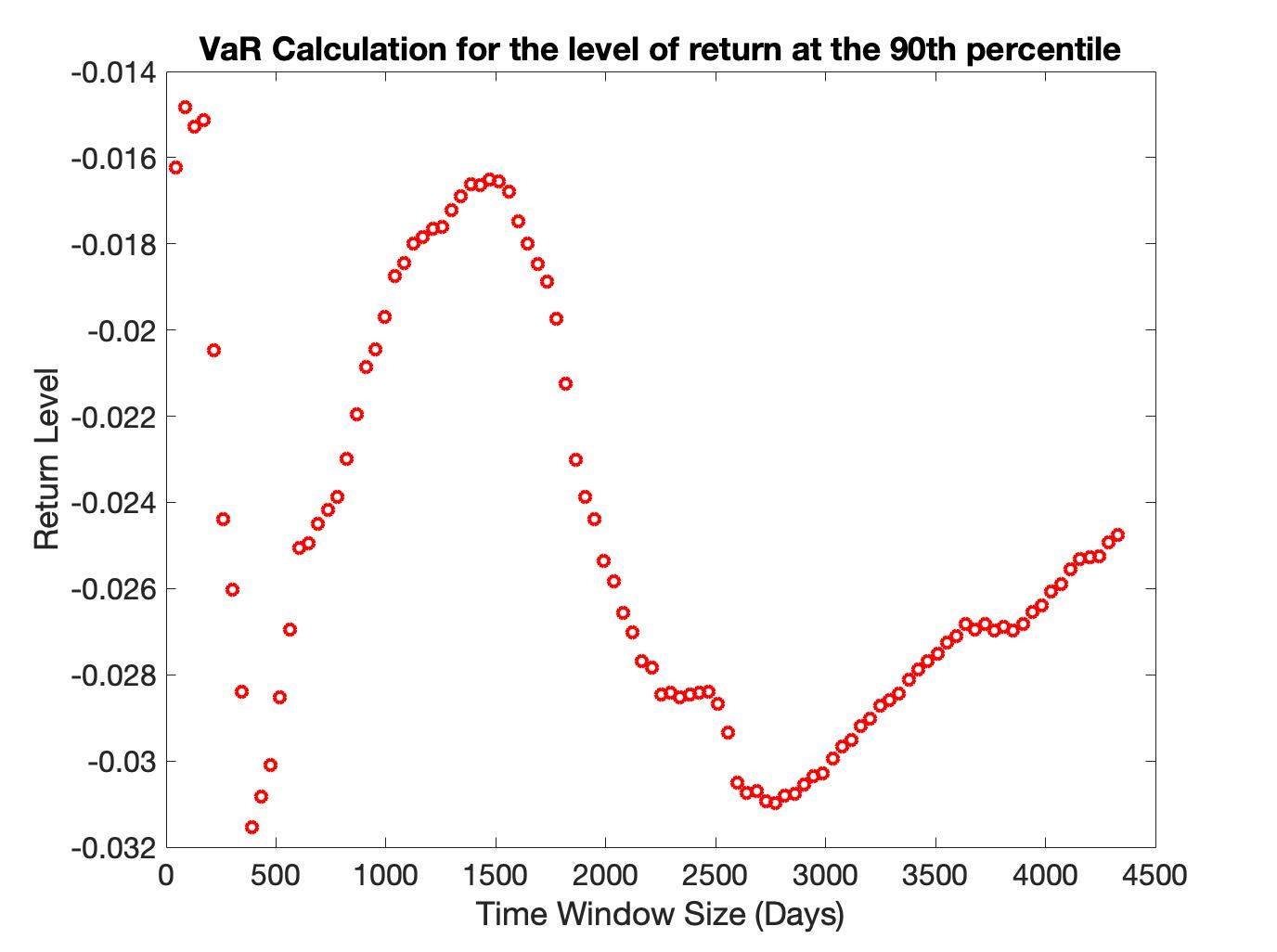}
            \caption[]%
            {{\small} Lloyds}    
            \label{lloyds_var}
        \end{subfigure}
        \hfill
        \begin{subfigure}[b]{0.475\textwidth}  
            \centering 
            \includegraphics[width=\textwidth]{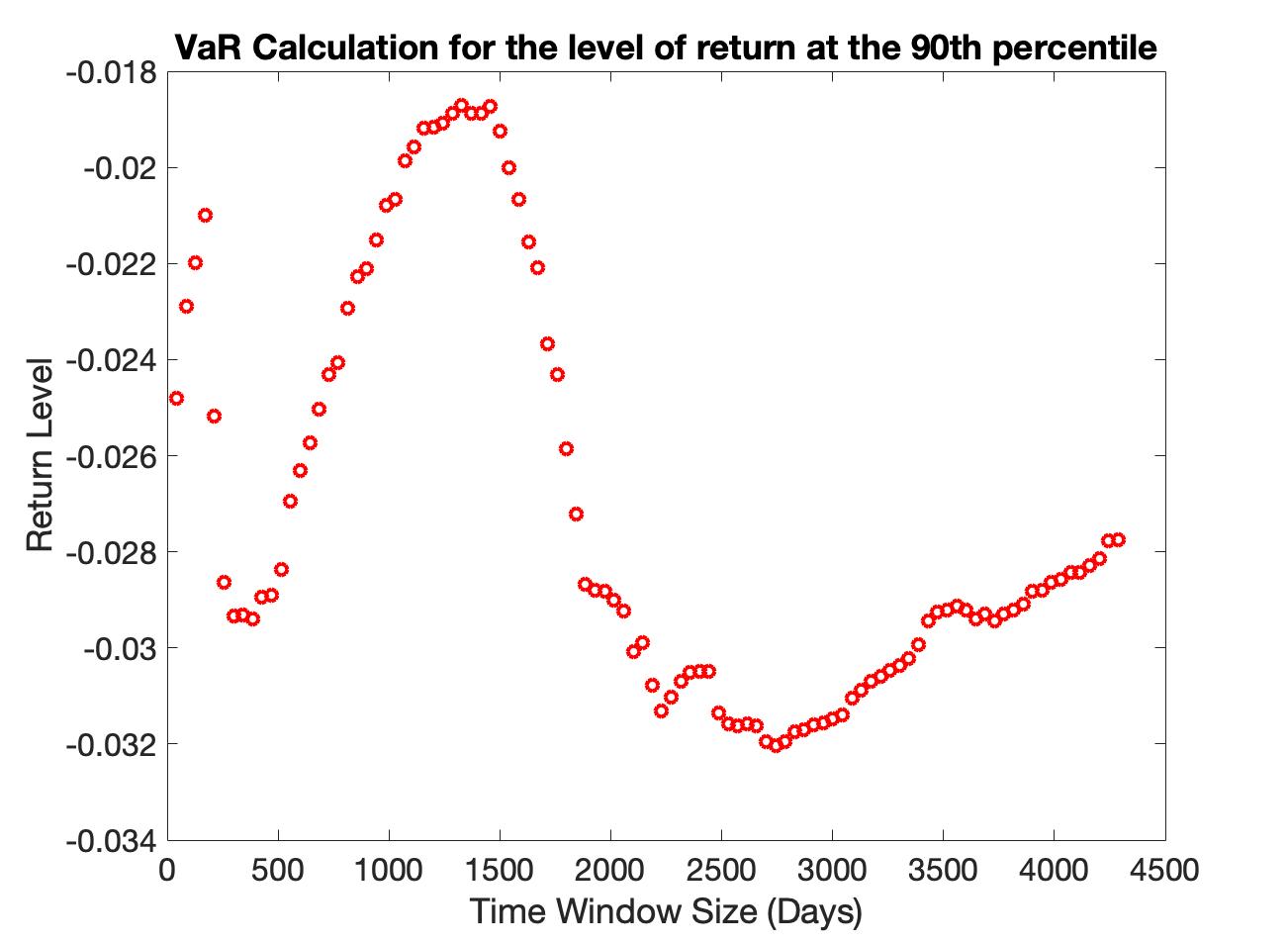}
            \caption[]%
            {{\small}Barclays}
            \label{barc_var}
        \end{subfigure}
        \vskip\baselineskip
        \begin{subfigure}[b]{0.475\textwidth}   
            \centering 
            \includegraphics[width=\textwidth]{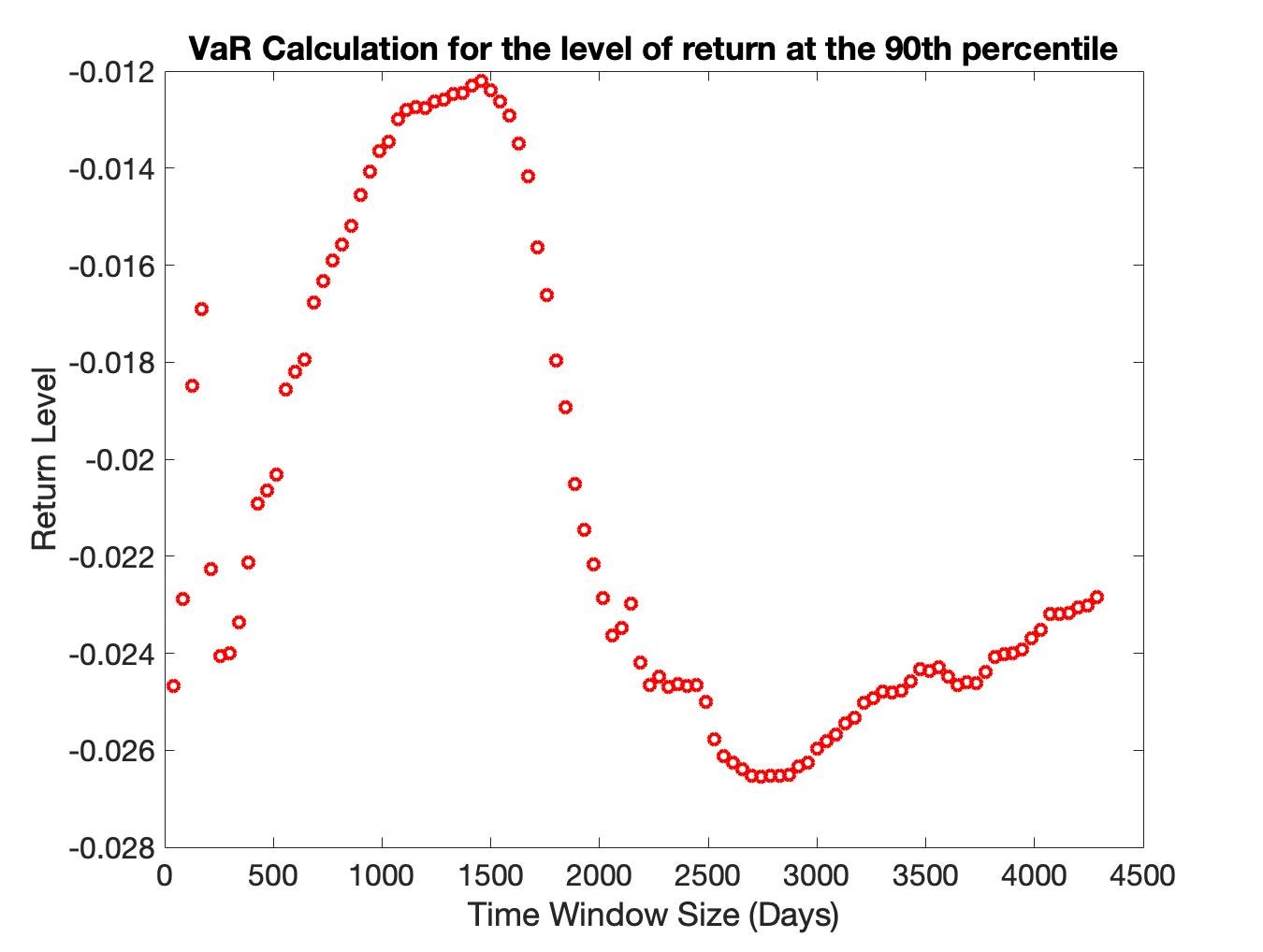}
            \caption[]%
            {{\small}Bank of America}    
            \label{boa_var}
        \end{subfigure}
        \hfill
        \begin{subfigure}[b]{0.475\textwidth}   
            \centering 
            \includegraphics[width=\textwidth]{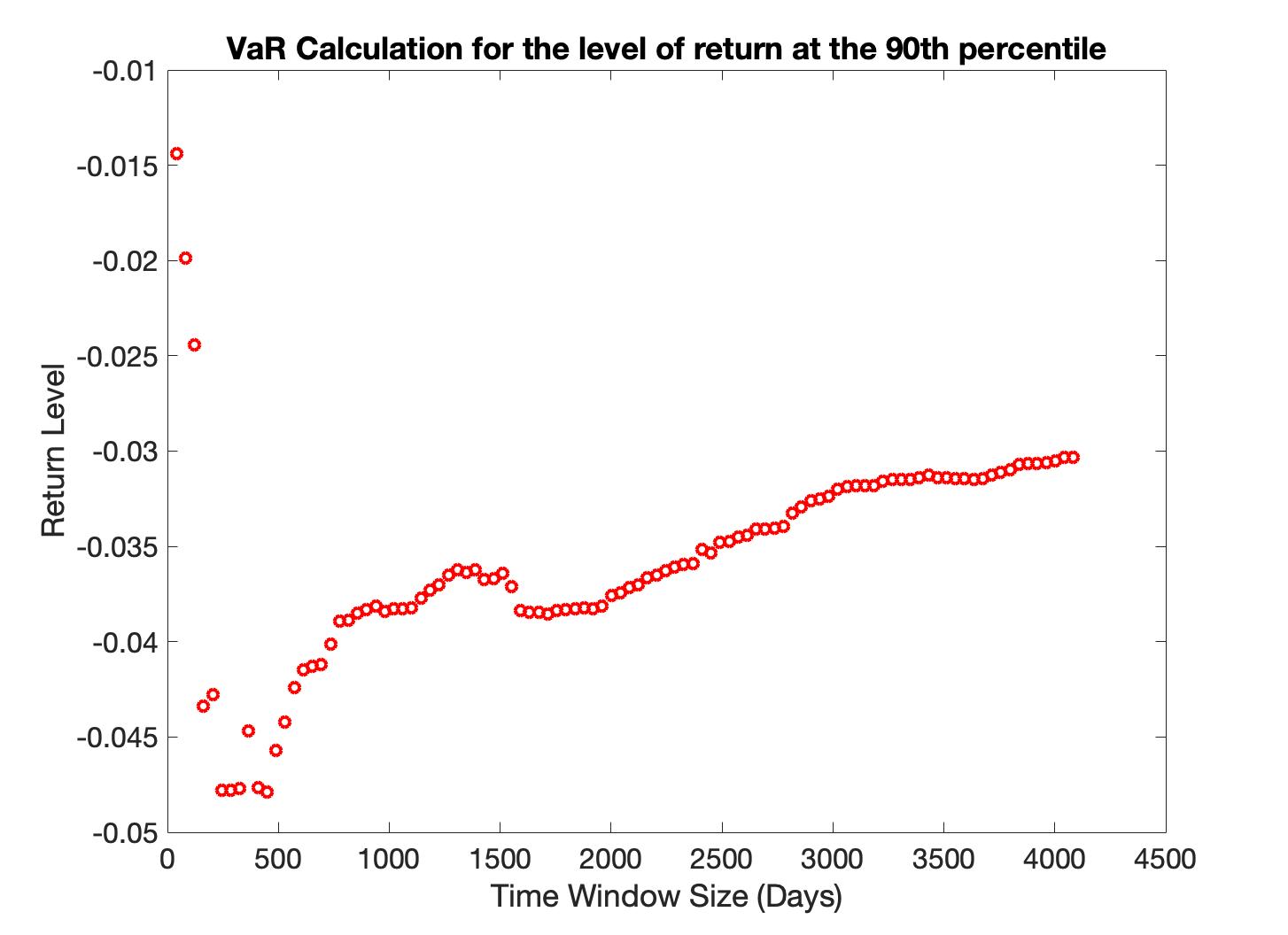}
            \caption[]%
            {{\small}Gold}    
            \label{gold_var}
        \end{subfigure}
        \caption[]
        {\small The Value-at-Risk Calculations for the 90th confidence level of returns for Lloyds, panel (a), Barclays, panel (b), Bank of America, panel (c) and Gold ETFs, panel (d). To construct such diagrams, we truncate the empirical data to $1\%$ of its original length and increment in 0.1$\%$ up to its total length. We then work out the 90th confidence level. We would expect to see a consistent increase in the level of loss of return in response to the increased time window. However, this is not seen and instead we see two distinct behaviours with regards to the level of loss. One region where we get sensical behaviour, increase losses for an increased time window and a second, where we get a decrease in the level of loss for the increased time window.} 
        \label{var}
 \end{figure*}

To deduce the implications of these different scaling relations on the level of risk, we use a simple Value at Risk (VaR) calculation, \cite{var1, var2}. We will analyse the level of loss that can be expected at the $90\%$ confidence interval for the corresponding time window. In figure \ref{var}, we see the level of return for the 90th confidence level for the truncated time windows for Lloyds Bank, Barclays bank, Bank of America and Gold ETFs.

In this figure, we can see some distinct regions in the level of risk. For the shortest time window, we have a relatively low level of loss, which is to be expected given the short time window. This equates to less uncertainty. When we increase the time window we get an increase in the loss, as we would expect, at the 90th confidence level. This consequently, is a result of the increase in the uncertainty in the return level given more data. However, starting from a certain point the increase stops and instead reverses. Now, we have a decreasing level of loss for an increasing time window. This is the point at which we gain a different scaling relation in the ($\Gamma_4$, $\Gamma_6$) phase space. However, there is now a disparity between the banking securities and gold. Whilst, the banking securities continue to decrease the level of loss with the time window, gold reverses again and starts to increase the loss for the increasing time window. For the banking securities, we get a much longer decrease with respect to the time window length. However, we do see this region end around the 200th data point and instead, we get a third regime where the level of loss starts to increase with increased length of time window, a behaviour we would expect.

\section{Hierarchical Analysis of Rare-Events}
In order to analyse the behavior of the higher order moments of the logarithm of price returns in the truncated time windows having $N$ trading days, we note that the rare-events whose probability is extremely low, will not contribute to the higher order moment calculation within this window. Indeed, the probability to observe tradings with returns $|x|>x_W$ occurring within $N$ days can be evaluated as $P_N(x_W)=2N\int_{x_W}^{\infty}p(x)dx$ with a probability distribution, $p(x)$, of the logarithm of price returns (for simplicity we assume $p(x)$ to be an even function). If $P\ll 1$, we can safely ignore such events and evaluate the higher order moments within interval $|x|<x_W$, where $x_W$ can be estimated from the condition that $P_N(x_W)=C\sim 1$, when $C$ is a constant. This allows us to evaluate the higher order moments for a $N$-days trading window, using the following equations:

\begin{eqnarray}
2N\int_{x_W}^{\infty}p(x)dx=C \nonumber \\
\langle x^n\rangle = 2\int_0^{x_W}x^n p(x)dx
\label{n-moments}
\end{eqnarray}
     
The empirically observed two distinct scaling laws suggest that the probability distribution should have two different functional behaviours at large $x$ resulting in a hierarchy of rare-events in two groups: rare-events and very rare-events. This can be done using an usual Paerto tail distribution whose exponent changes from $\gamma_1$ to $\gamma_2$ at certain $|x|=x_1$: 

\begin{equation}
 p(x) = 
  \begin{cases} 
   0 & \text{if } |x| < x_0 \\
   A \abs{\frac{x}{x_1}}^{-\gamma_1}       & \text{if } x_0 < |x| < x_1 \\
   A \abs{\frac{x}{x_1}}^{-\gamma_2}       & \text{if } |x| > x_1.
  \end{cases} 
  \label{paerto}
\end{equation}
Note that our analysis below does not depend on the behaviour of a probability density at small values of $|x|$, thus we assume that $p(x)=0$ for $|x|<x_0$, for simplicity of our estimations. Substituting equation (\ref{paerto}) into the set of equations (\ref{n-moments}) and restricting our analysis to $3<\gamma_1<5$ and $3<\gamma_2<5$, we derive the short-window scaling relations for $x_0\ll x_W\ll x_1$:

\begin{eqnarray}
x_W=R_1N^{\frac{1}{1-\gamma_1}}, \nonumber \\
\Gamma_4=K_4 (x_W)^{5-\gamma_1}, \nonumber \\
\Gamma_6=K_6 (x_W)^{7-\gamma_1}
\label{scaling-1}
\end{eqnarray}
and the long-window scaling relations for $x_1\ll x_W$:

\begin{eqnarray}
x_W=R_2N^{\frac{1}{1-\gamma_2}} \nonumber \\
\Gamma_4=Q_4 (x_W)^{5-\gamma_2}, \nonumber \\
\Gamma_6=Q_6 (x_W)^{7-\gamma_2}
\label{scaling-2}
\end{eqnarray} 
where $R_1, R_2, K_4, K_6, Q_4, Q_6$ do not depend on $N$.
When deriving the above equation we keep only the main contributions to the integrals, for example, approximating $(x_0)^{3-\gamma_1}-(x_W)^{3-\gamma_1}\approx (x_0)^{3-\gamma_1}$ and $(x_W)^{5-\gamma_1}-(x_0)^{5-\gamma_1}\approx (x_W)^{5-\gamma_1}$.

From the above set of equations we derive two different scaling laws, these laws have been observed in the empirical data:

\begin{equation}
\Gamma_ 6 = L_1 \Gamma_4 ^{\frac{7-\gamma_1}{5-\gamma_1}}
\end{equation}
which is valid for short time windows $N\ll (x_1/R_1)^{\gamma_1-1}$, and:

\begin{equation}
\Gamma_6 = L_2 \Gamma_4^{\frac{7-\gamma_2}{5-\gamma_2}},
\end{equation} 
which is valid for long time windows, $N\gg (x_1/R_2)^{\gamma_2-1}$. Where, $L_1$ and $L_2$ are constants. In order to reproduce the empirical observation that the shorter time windows have a steeper gradient in the ($ln(\Gamma_4)$, $ln(\Gamma_6)$) space we have to request that $\gamma_1 > \gamma_2$. This means, the steeper the gradient in this higher order moment space, equates to a faster decay in the probability distribution with respect to price change, $x$. Using this analysis, we are able to see that depending where we take this truncation, $x_W$, we will potentially expose ourselves to a higher level of risk to these rare-events. This is due to the truncation determining the forecasting horizon we are interested in. Therefore, we can see that the length of window we wish to model will have an effect on the risk level we expose ourselves to during this period. In order to hedge for a higher level of risk than we may expect, we must ascertain which exponent of the power law the distribution has for events occurring with frequency $1/N$, when forecasting a $N$ trading day horizon. In doing so, we can correctly calculate the risk we are exposing our position to.

\section{Conclusion}
\label{conc}
By the use of higher order moments, we uncover a new scaling behaviour of the empirical data. For the longer time windows, the logarithm of the fourth and sixth order standardised moments follow a straight line. The same behaviour was observed for the shorter time windows but with different parameters of the scaling equation. This fact is seen throughout all of the empirical data we have analysed for different financial data series.

We also highlight the impact of differing economic periods upon these scaling relations via the investigation of the empirical data throughout the 2008 financial crash. Here, we show that for companies directly affected by the crash, primarily banking companies, there is a drastic change to the logarithmic gradient of the scaling relation. This impact is long lasting in the empirical data. Almost a decade after the crash there is still an evident legacy of this economic period in the empirical data's higher order moments.

Furthering the investigation of empirical data, we show the relationship between the higher order standardised moments of the empirical data and the standardised moment values of the gaussian distribution. We show by truncating the data into 18 years, 3 years and a 6 month time series, the length of time we investigate over has a stark impact on the higher order standardised moments.

Moreover, we try to replicate the phenomena using a GARCH-double-normal(1,1) model. We are able to show that for all parameter values investigated, we gain two distinct scaling relations. However, we get the longer time window's scaling relation to have a larger logarithmic gradient, $B$, than the shorter one. A clear contradiction to the empirical data. We resolve this rather puzzling behaviour by modelling rare-events in different time windows.

In order to deduce the behaviour of risk we carry out a Value-at-Risk type calculation to determine the potential loss at the 90th confidence level for the different time horizons analysed. We see that for the different scaling relations we encounter different levels of risk. We would expect that for an increasing time horizon, the risk increases, however, when we have the observed change in scaling relation we encounter a reduction in the level of loss for an increase in time horizon. In conclusion, the data analysis reported in this paper can elucidate the behaviour of prices within short and long time horizons and as such can be used as a useful tool for market and portfolio analysis. 

\newpage

\section{References}

\bibliographystyle{IEEEtran}
\bibliography{empirical_database}

\appendix

\newpage

\section{Standardised Moments Order and Gaussian Values}
\label{kurt}

\begin{table}[h!]
\centering
\begin{tabular}[c]{|p{2cm}|p{2cm}|p{2cm}|}
\hline
Calculation Number&Equation&Gaussian Value, $\Gamma_n^{gaussian}$\\ \hline
A&$\frac{\langle (x-\mu)^4 \rangle}{\langle (x-\mu)^2 \rangle^2}$&3\\ \hline
B&$\frac{\langle (x-\mu)^6\rangle}{\langle (x-\mu)^2\rangle^3}$&15\\ \hline
C&$\frac{\langle (x-\mu)^8\rangle}{\langle (x-\mu)^2\rangle^4}$&105\\ \hline
D&$\frac{\langle (x-\mu)^{10}\rangle}{\langle (x-\mu)^2\rangle^5}$&945\\ \hline
E&$\frac{\langle (x-\mu)^{12}\rangle}{\langle (x-\mu)^2\rangle^6}$&10395\\ \hline
\end{tabular}
\caption{The equations of the standardised moment calculations for those used in figure \ref{kurt_ratios}. We also display the corresponding gaussian values.}
\label{kurt_gauss}
\end{table}

\begin{table}[h!]
\centering
\begin{tabular}[c]{|p{2cm}|p{2cm}|p{2cm}|p{2cm}|p{2cm}|p{2cm}|}
\hline
Standardised Moment Calculation Number&Bank of America&Barclays Bank&Citigroup&HSBC&Lloyds Bank\\ \hline
A&16.51&46.47&7.14&6.07&26.02\\ \hline
B&1.46e03&1.18e04&124.17&98.55&2.58e03\\ \hline
C&1.78e05&3.37e06&3.12e03&2.65e03&3.22e05\\ \hline
D&2.30e07&9.65e08&9.32e04&9.15e04&4.35e07\\ \hline
E&3.04e09&2.77e11&3.05e06&3.56e06&6.16e09\\ \hline
\end{tabular}
\caption{The values of the raw data standardised moments for the 18 year time series for the companies used in figures \ref{kurt_ratios}, \ref{short_kurt}. The ordering of the standardised moment calculations are the same as those seen in table \ref{kurt_gauss}.}
\label{1}
\end{table}

\begin{table}[h!]
\centering
\begin{tabular}[c]{|p{2cm}|p{1.7cm}|p{1.7cm}|p{1.7cm}|p{1.7cm}|p{1.7cm}|p{1.7cm}|}
\hline
Standardised Moment Calculation & Barclays & BoA    & Citi   & Lloyds & GSK      & DowJones \\  \hline
A & 4.4013   & 4.9296 & 4.1463 & 3.369  & 6.1169   & 8.6692   \\  \hline
B & 36.017   & 41.762 & 29.324 & 17.696 & 79.605   & 149.98   \\  \hline
C & 389.98   & 449.33 & 264.94 & 114.97 & 1391.8   & 3274.5   \\  \hline
D & 4864.9   & 5442.7 & 2730.9 & 845.15 & 27453    & 79051    \\  \hline
E & 65468    & 69803  & 30472  & 6733.8 & 5.69E+05 & 2.00E+06 \\  \hline
\end{tabular}
\caption{The values of the raw data of the standardised moment calculations for the three year time series of the companies used in figure \ref{short_kurt}, specifically figure \ref{3year}, the ordering of the values is the same as that shown in table \ref{kurt_gauss}.}
\label{kurt_short_table}
\end{table}

\newpage

\section{The fourth and eighth order standardised moments}
\label{g8_scale}

\begin{figure}[h!]
\centering
\begin{subfigure}{0.5\linewidth}
  \centering
  \includegraphics[width=\linewidth]{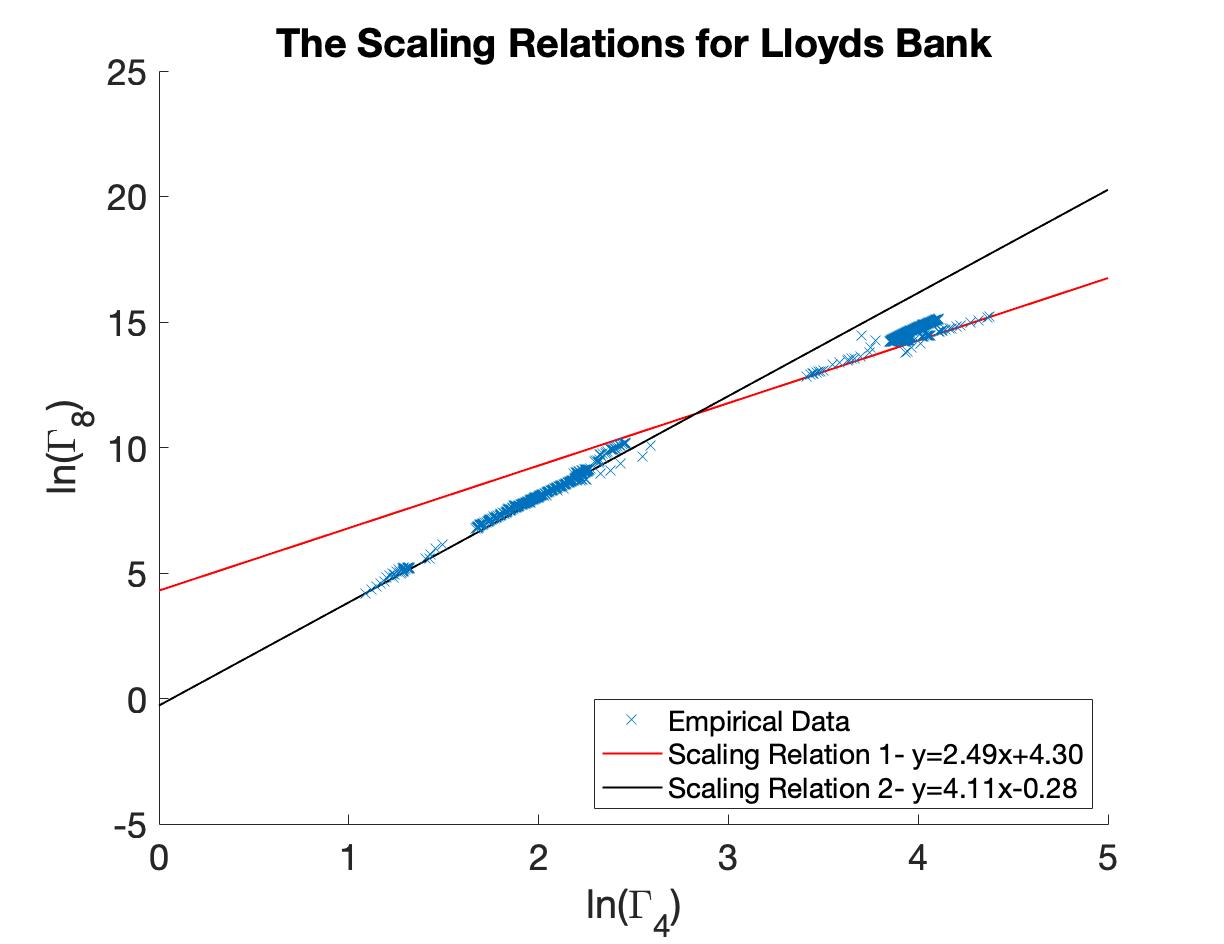}
  \caption{}
  \label{}
\end{subfigure}%
\begin{subfigure}{0.5\linewidth}
  \centering
  \includegraphics[width=\linewidth]{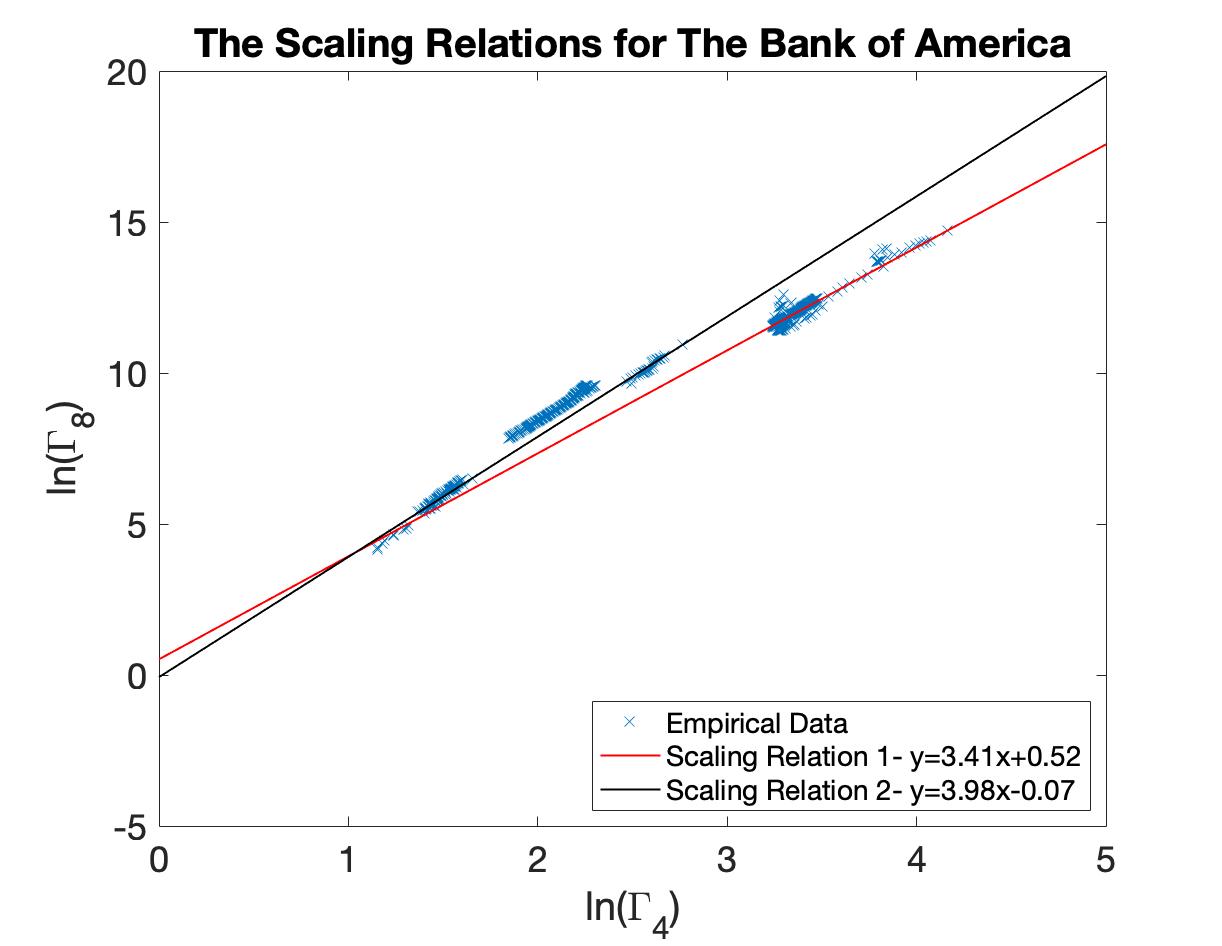}
  \caption{}
  \label{}
\end{subfigure}
\caption{Here, we present the scaling relations in logarithmic space for the fourth and eighth order standardised moments. In panel (a), we show the relation for Lloyds Bank, with the shorter scaling relation being, $y=4.11x-0.28$, shown in black and the longer relation, $y=2.49x+4.30$, in red. In panel (b), we see the scaling relations for Bank of America, we see for the shorter time windows, the scaling relation is, $y = 3.98x-0.07$, in black, whilst the longer time windows have a scaling relation of, $y = 3.41x+0.52$, in red.}
\label{}
\end{figure}

\end{document}